\documentclass{article}
\usepackage{amsmath}
\usepackage{pifont}
\usepackage{spconf}
\usepackage{longtable}
\usepackage{rotating}
\usepackage{multirow}
\usepackage{graphicx}
\usepackage{subfig}
\usepackage{enumerate}
\usepackage{indentfirst}
\usepackage{amsthm}

\usepackage{graphics}
\usepackage{epsfig}
\usepackage{bm}
\usepackage{epstopdf}
\usepackage{amssymb}
\usepackage{amsfonts }
\usepackage{graphicx}
\usepackage{booktabs}
\usepackage{float}
\usepackage[ruled,linesnumbered,boxed]{algorithm2e} 
\usepackage[marginal]{footmisc}
\usepackage{color}
\usepackage{cite}
\usepackage{array}
\usepackage{tabularx}

\usepackage{makecell}
\usepackage{setspace}
\usepackage{hyperref}
\usepackage{bibspacing}
\hypersetup{hypertex=true,
	colorlinks=true,
	linkcolor=blue,
	anchorcolor=blue,
	citecolor=blue}

\begin{document}\sloppy

\def\x{{\mathbf x}}
\def\L{{\cal L}}

\title{GA-HQS: MRI reconstruction via a generically accelerated unfolding approach}
%
\name{Jiawei Jiang, Yuchao Feng, Honghui Xu, Wanjun Chen, Jianwei Zheng}
\address{Zhejiang University of Technology}

\maketitle

\begin{abstract}
Deep unfolding networks (DUNs) are the foremost methods in the realm of compressed sensing MRI, as they can employ learnable networks to facilitate interpretable forward-inference operators. However, several daunting issues still exist, including the heavy dependency on the first-order optimization algorithms, the insufficient information fusion mechanisms, and the limitation of capturing long-range relationships. To address the issues, we propose a Generically Accelerated Half-Quadratic Splitting (GA-HQS) algorithm that incorporates second-order gradient information and pyramid attention modules for the delicate fusion of inputs at the pixel level. Moreover, a multi-scale split transformer is also designed to enhance the global feature representation. Comprehensive experiments demonstrate that our method surpasses previous ones on single-coil MRI acceleration tasks.
\end{abstract}
\begin{keywords}
Compressed sensing MRI, accelerated unfolding, half-quadratic splitting, information fusion
\end{keywords}
\section{Introduction}
Magnetic resonance imaging (MRI) is widely deployed in clinical medical diagnosis due to its non-invasive ability of capturing abundant biological tissues. However, the prolonged scan time of MRI in k-space probably causes motion artifacts and patient discomfort, which hinders the further application for time-critical diagnoses, such as stroke. One option for accelerating MRI is to conduct significant undersampling in k-space, followed by an algorithmic reconstruction, i.e., compressed sensing (CS). Nevertheless, due to the violation of Nyquist sampling theorem, the k-space undersampling often leads to aliasing artifacts in the image domain. Therefore, a delicate design of the reconstruction method is crucial for obtaining high-quality MR images. \par
To solve the highly ill-posed inverse problem, on the one hand, traditional methods, such as total variation (TV) \cite{block2007undersampled} and wavelets \cite{qu2012undersampled}, have been widely investigated \cite{lustig2007sparse,liang2009accelerating}. These solutions incrementally enhance the quality of each image through iterations, often costing more running time and leading to over-smoothed recoveries. Besides, the requirement of a fine tuning for the multiple hyper-parameters is also a daunting issue. On the other hand, in recent years, deep neural networks (DNNs) have made breakthroughs in visual inverse problems. The early research of DNNs on MRI reconstruction focus on utilizing some existing CNN architectures, such as Unet\cite{zbontar2018fastmri} and ResNet\cite{lee2018deep}, to directly invert the forward degradation. Such schemes, although with acceptable quality and fast inference, often lack the rational interpretability and may increase the risk of instability. As an alternative, deep unfolding network (DUN) is a promising orientation for accelerating MRI due to its well interpretability and outstanding outcome. Technically, these approaches first model MRI acceleration as an optimization problem, and subsequently achieve DUN by replacing certain components in the iterations with learnable network blocks consisting of a fixed number of stages \cite{yang2018admm,zhang2018ista,aggarwal2018modl,luo2022generalized}. For example, based on the traditional model-based CS method, ADMM-CSNet \cite{yang2018admm} adopts convolutional layer to approximate the sparsity term in the ADMM framework and achieves favorable reconstruction accuracy. ISTA-Net \cite{zhang2018ista} unfolds the update-steps of conventional iterative shrinkage-threshold algorithm (ISTA) into a network, with each stage corresponding to a single iteration of ISTA. MoDL \cite{aggarwal2018modl} unrolls an alternating recursive algorithm consisting of interleaved CNN blocks and fidelity terms solved by conjugate gradient (CG) optimization. \par

Although remarkable achievements have been made, following an in-depth investigation, one can see that the potential of DUNs has not been fully explored. First of all, existing DUNs are mainly based on the typical first-order optimization algorithms. To the best of our knowledge, the role of second-order acceleration algorithm has not been reported. Second, although the specific forms of different algorithms vary greatly, the purpose of the practical operations can be summarized into two types, namely, \textbf{information fusion} (\textbf{IF}) and \textbf{image calibration} (\textbf{IC}). Given multiple inputs, most previous works have dealt with IF issue in a simple arithmetic way, whose data flow is usually calculated using the original formula or simply imposing a re-weighting operation. Such coarse interaction not only fails to effectively fuse the input information, but also limits the further feature enhancement. As for the IC issue, a neural network is often employed to learn the potential denoising prior. However, the existing approaches depend purely on CNNs, suffering from the thirst for establishing non-local similarities and long-range dependencies, both of which are critical for fast MRI reconstruction.\par
\begin{figure*}[htbp]
	\centering
	\includegraphics[width=\linewidth,scale=1.00]{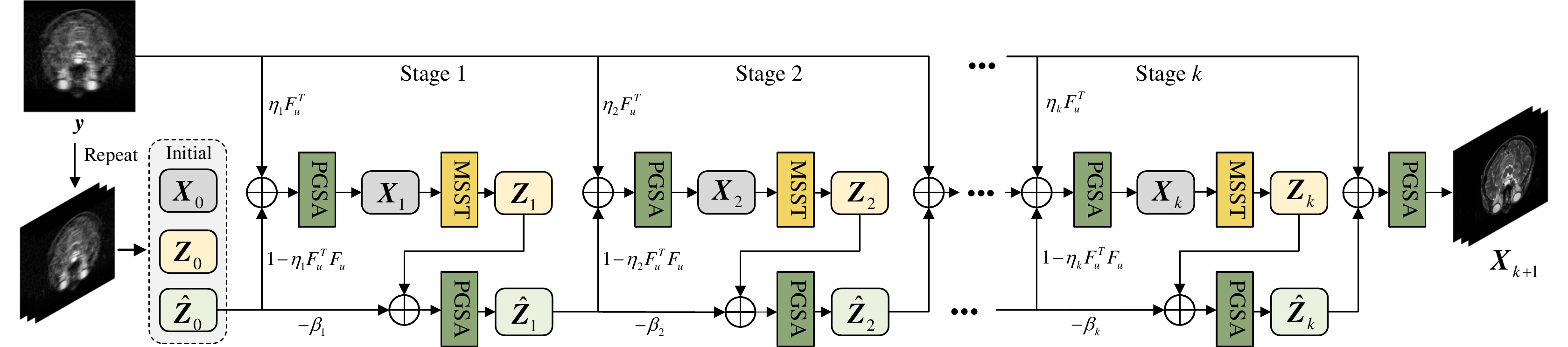}
	\centering
	\caption{The architecture of GA-HQS with $k$+1 stages. The input data $\bm{X}_0,\bm{Z}_0,\hat{\bm{Z}}_0$ is the concatenation of $m$ replicas of the complex-valued zero-filled images $\bm{x}_0$. In accordance with  Eq. \ref{GA-hqs}, each stage primarily consists of three computing nodes, the information fusion node $\bm{X}_k$, the image calibration node $\bm{Z}_k$, and the acceleration node $\hat{\bm{Z}}_k$.    }
	\label{fig:framework}
	\vspace{-1pt}
\end{figure*}
To address the above problems, we first extend a widely used first-order algorithm, i.e.,  Half-Quadratic Splitting (HQS), to an accelerated version for fast MRI reconstruction, dubbed as A-HQS. Such accelerating scheme is essentially a second-order operator and has been extensively investigated and proven to be effective \cite{franca2018admm,tang2022accelerating}. Second, in the information communication phase, a novel pyramid gated-squeeze attention (PGSA) module is designed as the fuser. Functionally, PGSA extracts hierarchical multi-scale spatial information and constructs feature-gated channel attention, in which group convolution is introduced to control the computational cost against the increasing of different kernel scales. Finally, in order to extract global similarities and long-range dependencies, we customize a multi-scale split transformer (MSST) as the denoiser in image calibration step, which mainly contains two branches, i.e., split depth-wise convolution (SDC) and half-shuffle attention Unet (HSAU). SDC maps input to features with multi-scale receptive fields, while HSAU refines the local contextual information and adaptively aggregates the global dependencies. We plug PGSA and MSST into A-HQS to build an iterative architecture, namely a generically accelerated half-quadratic splitting (GA-HQS). With the proposed components, GA-HQS significantly outperforms other state-of-the-art (SOTA) deep unfolding algorithms by 2dB on single-coil MRI reconstruction, and even outperforms some multi-modal methods on several tasks (e.g., 4x acceleration on IXI datasets under radial mask). The framework of GA-HQS is depicted in Fig. \ref{fig:framework}. \par
	\vspace{-10pt}

\section{Proposed method}
\subsection{ Problem Formulation}
The MRI imaging systems are considered to reconstruct the recovery image $\bm{x} \in \mathbb{C}^{H\times W}$ from the k-space undersampled measurement $\bm{y}\in \mathbb{C}^{H\times W}$ as
\begin{align}\label{eq1}
	\min \limits_{\bm{x}} \frac{1}{2} \Vert \bm{y}-F_u\bm{x}\Vert^2_2 + \lambda \mathcal{R}(\bm{x}),
\end{align}
When MRI is single-coil, $F_u$ can be obtained by multiplying the binary undersampling mask $M$ and the Fourier transform $F$, i.e, $F_u=MF$. In the case of multi-coil, $F_u = MF\mathcal{P}$, where $\mathcal{P}$ is the coil sensitivity. In this study, we focus on the issue of single coil reconstruction. $\mathcal{R}(\bm{x})$ denotes the regularization term on $\bm{x}$ and $\lambda$ is a parameter balancing the trade-off between the two terms.

\subsection{GA-HQS Unfolding Framework}
For MRI reconstruction, previous unfolding frameworks ignored the role of accelerated optimization algorithms. More precisely, the second-order gradient information is not fully utilized. To build a bridge between the two sides, we extend the widely used HQS algorithm in DUN to an accelerated version. For ease of analysis, we will directly give the iterative formulation of HQS and A-HQS for solving the problem \eqref{eq1}. The update steps of HQS is as follows:
\begin{subequations}\label{hqs}
	\begin{align} \label{hqs-a}
		&\bm{x}_{k+1} = \arg\min\limits_{\bm{x}} \frac{1}{2} \Vert \bm{y}-F_u\bm{x}\Vert^2_2 + \frac{\mu_{k+1}}{2}\Vert \bm{x}-\bm{z}_k\Vert^2_2,  \\ \label{hqs-b}
		&\bm{z}_{k+1} = \textbf{proxNet}_{\mathcal{R},\tau} \left( \bm{x}_{k+1}\right).
 	\end{align}
\end{subequations}
The update steps of the extended version, i.e., A-HQS, is as follows:
\begin{subequations}\label{a-hqs}
	\begin{align} \label{a-hqs-a}
		&\bm{x}_{k+1} = \arg\min\limits_{\bm{x}} \frac{1}{2} \Vert \bm{y}-F_u\bm{x}\Vert^2_2 + \frac{\mu_{k+1}}{2}\Vert \bm{x}-\hat{\bm{z}}_k\Vert^2_2 , \\ \label{a-hqs-b}
		&\bm{z}_{k+1} = \textbf{proxNet}_{\mathcal{R},\tau} \left( \bm{x}_{k+1}\right), \\   \label{a-hqs-c}
		&\hat{\bm{z}}_{k+1} = \bm{z}_{k+1} + \beta_{k+1}(\bm{z}_{k+1} -\bm{z}_{k}).
	\end{align}
\end{subequations}
where $ \mu_{k+1} $ is a penalty parameter and $\textbf{proxNet}_{\mathcal{R},\tau}(\bm{u})= \arg\min_{\bm{v}}\frac{1}{2}\Vert \bm{v}-\bm{u} \Vert_2^2 + \tau g(\bm{v})$. Each update in HQS has two steps $\bm{x}_{k+1}$ and $\bm{z}_{k+1}$, while each in A-HQS has one extra acceleration step $\hat{\bm{z}}_{k+1}$. To convert them into networks, the formulations of $\bm{x}_{k+1}$ in \eqref{hqs-a} and \eqref{a-hqs-a}, $\hat{\bm{z}}_{k+1}$ in \eqref{a-hqs-c} are generally keep unchanged, with $ \mu_{k+1}$ and $\beta_{k+1}$ set to trainable parameters. $\textbf{proxNet}_{\mathcal{R},\tau}(\bm{u})$ can be regarded as a denoising process with $\bm{x}_{k+1}$ as input \cite{dong2018denoising}. Therefore, a denoising network $\mathcal{D}_k$ is usually adopted to solve Eq. \eqref{hqs-b} and \eqref{a-hqs-b}. Obviously, the main difference between HQS and A-HQS is the acceleration variable $\hat{\bm{z}}_{k+1}$, which is a second-order term in the framework of ordinary differential equations \cite{su2014differential} and can be represented by a skip connection in the neural network architecture. Next, we will focus on A-HQS for further analysis. Note \eqref{a-hqs-a} involves a quadratic regularized least-squares problem, thus it has a closed-form solution written as
\begin{figure}[tbp]
	\centering
	\includegraphics[width=\linewidth,scale=1.00]{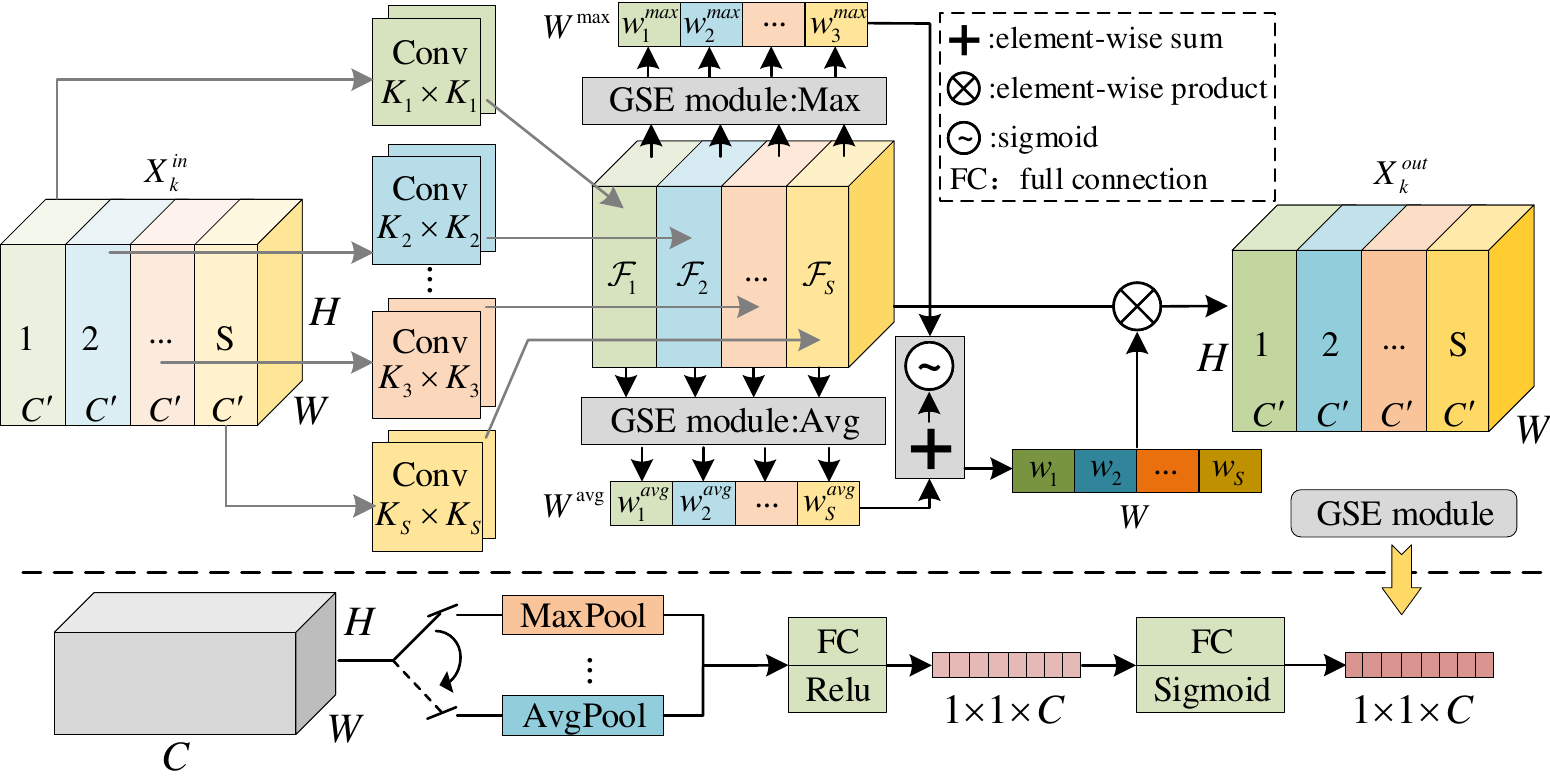}
	\centering
	\caption{Diagram of PGSA. PGSA divides the input $\bm{X}_k^{in}$ into $S$ blocks, each of which is mapped by $S$ multi-scale convolutions, followed by a gating mechanism to adaptively compute channel attentions. }
	\label{fig:channel}
	\vspace{-1pt}
\end{figure}
\begin{align}
 \bm{x}_{k+1} =  (F_u^{T}F_u + \mu \bm{I})^{-1}(F_u^{T} \bm{y} + \mu \hat{\bm{z}}_{k}),
\end{align}
where $\bm{I}$ is an identity matrix. Since the dimensionality of $(F_u^{T}F_u + \mu \bm{I})$ will be large, we empoly the Sherman–Morrison–Woodbury (SMW) matrix inversion \cite{afonso2010fast} formula to simplify the computation. Based on $F_uF_u^{T} = \bm{I}$, it follows
\begin{align}
	\bm{x}_{k+1} =  \hat{\bm{z}}_{k} + \eta F_u^{T}(\bm{y}-F_u\hat{\bm{z}}_{k} ),
\end{align}
where $\eta = \frac{1}{\mu+1}$. Then we formulate the iterative scheme of A-HQS as:
\begin{subequations}\label{D-AHQS}
	\begin{align}  \label{f_a}
		&\bm{x}_{k+1} =  \hat{\bm{z}}_{k} + \eta F_u^{T}\bm{y}-\eta F_u^{T}F_u\hat{\bm{z}}_{k},  \\
		&\bm{z}_{k+1} = \mathcal{D}_k(\bm{x}_{k+1}) ,\\ \label{f_c}
		&\hat{\bm{z}}_{k+1} = (1+ \beta_{k+1})\bm{z}_{k+1} -\beta_{k+1} \bm{z}_{k}.
	\end{align}
\end{subequations}
As previously mentioned, it can be observed that each stage is actually composed of two modules: information fusion (\textbf{IF}) for $\bm{x}_{k+1}$ and $\hat{\bm{z}}_{k+1}$, image calibration (\textbf{IC}) for $\bm{z}_{k+1}$. \par
Previous DUNs have devoted to investigating different denoising operators $\mathcal{D}_k$ of IC module to break the bottleneck, while neglecting to explore more effective strategies for information fusion. In this work, we argue that the pixel-level summation of different terms in \eqref{f_a} and \eqref{f_c} may be coarse, which greatly limits the information transmission. Therefore, we propose a fine modulated HQS (GA-HQS) to implement a more effective information fusion. The iterative scheme of our GA-HQS are as follows:

\begin{figure}[tbp]
	\centering
	\includegraphics[width=\linewidth,scale=1.00]{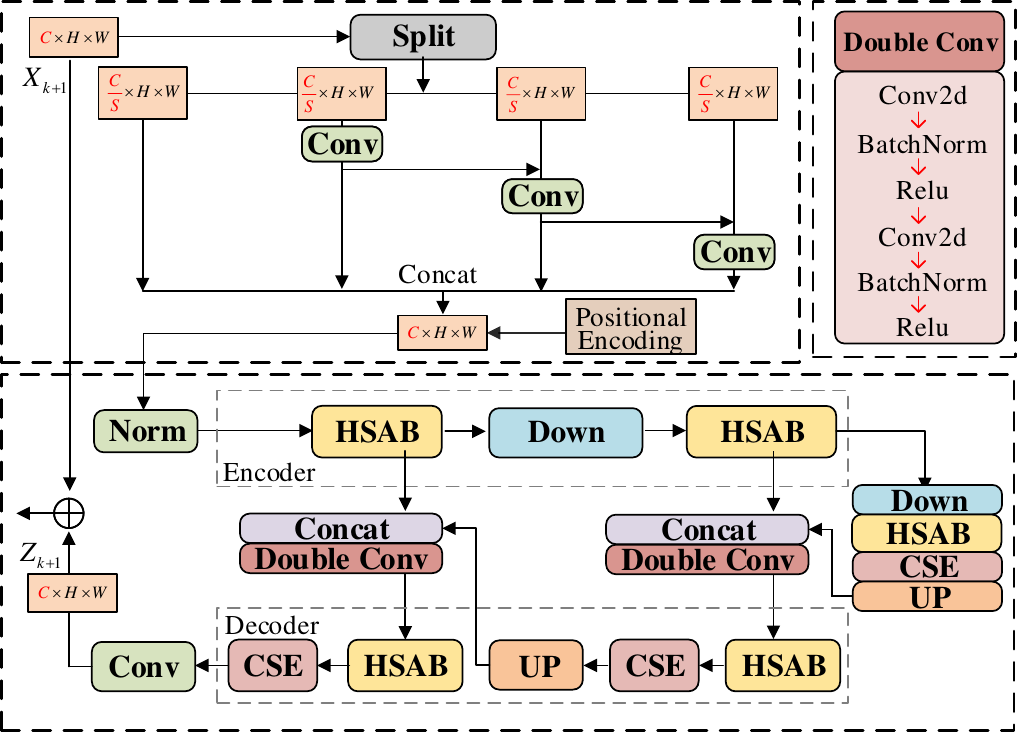}
	\centering
	\caption{Diagram of MSST. MSST is composed of two branches: split-depth-wise convolution (SDC) and half-shuffle attention Unet (HSAU). The former adopts the same splitting mechanism as PGSA to expand the receptive field, while the latter is composed of modules such as HSAB and CSE to capture global dependencies.}
	\label{fig:unet}
	\vspace{-10pt}
\end{figure}
\begin{subequations}   \label{GA-hqs}
	\begin{align}  \label{fm-hqsa}
		&\bm{X}_{k+1}: \bm{x}_{k+1} =  \mathcal{C}_{att} \left(\hat{\bm{z}}_{k} \oplus \eta F_u^{T}\bm{y} \oplus (-\eta F_u^{T}F_u)\hat{\bm{z}_{k}} \right), \\
		&\bm{Z}_{k+1}: \bm{z}_{k+1} = \mathcal{D}_k(\bm{x}_{k+1}), \\  \label{fm-hqsc}
		&\hat{\bm{Z}}_{k+1}: \hat{\bm{z}}_{k+1} = \mathcal{C}_{att} \left((1\!+\! \beta_{k+1})\bm{z}_{k+1} \oplus (-\beta_{k+1} \bm{z}_{k}) \right).
	\end{align}
\end{subequations}
where $\oplus$ is the concat operator. Instead of the summation operator used in \eqref{D-AHQS}, we concatenate the components of \eqref{fm-hqsa} and \eqref{fm-hqsc}, respectively, followed by a channel attention operator $\mathcal{C}_{att}$. With the aid of attention mechanism, such alterations increase the flexibility of information transmission. Next, the detailed implementation of GA-HQS would be given.\par

\begin{figure*}[htbp]
	\centering
	\includegraphics[width=\linewidth,scale=1.00]{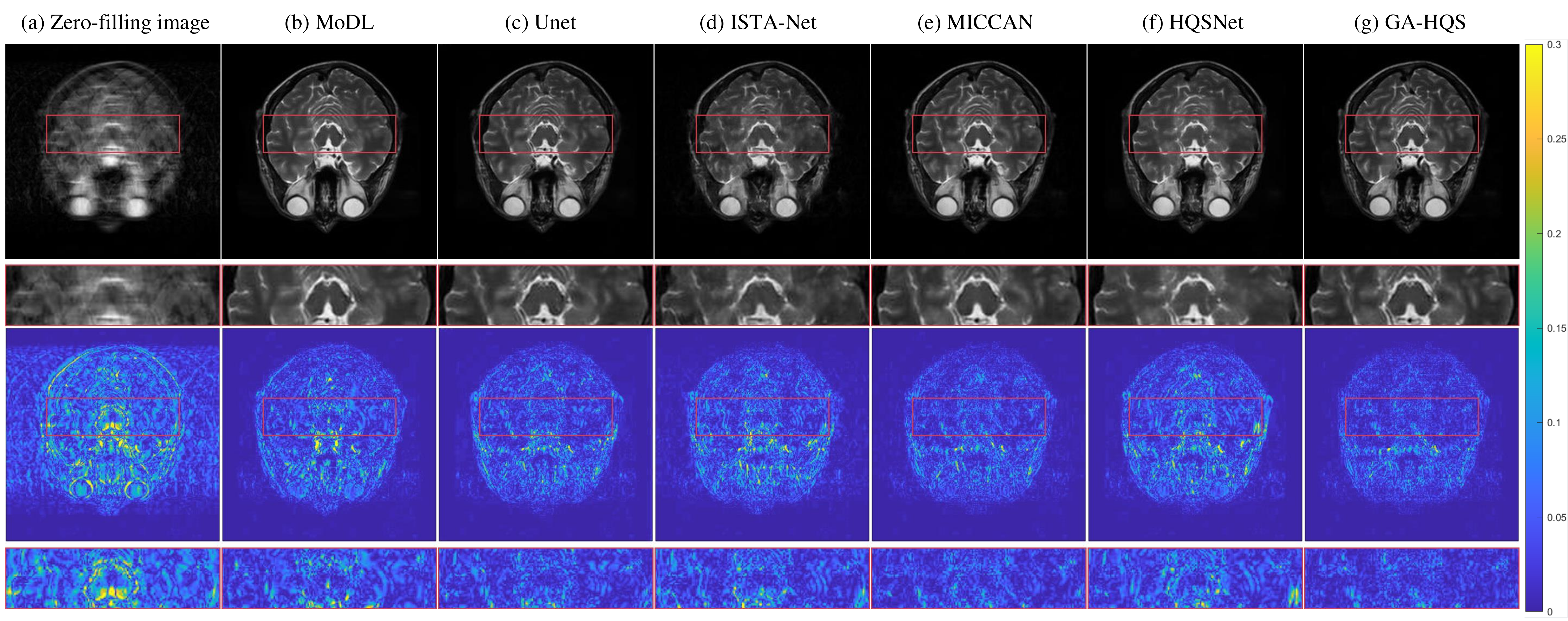}
	\centering
	\caption{Visualization of reconstruction result and corresponding error map with 8$\times$ Acceleration under \textit{equispaced fraction} mask for IXI dataset. The red boxed area is enlarged to show details.}
	\label{fig:visual1}
	\vspace{-1pt}
\end{figure*}
\subsection{Network Architecture}
\textbf{IF module}: The channel attention mechanism allows adaptively adjusting the weight of each channel and producing a more informative output, which is naturally suited to our IF module. Therefore, we design a Pyramid Gated-Squeeze Attention (PGSA) module to play the role of $\mathcal{C}_{att}$ and implement  fine-grained  feature fusion along with the channels.\par

As shown in Fig. \ref{fig:channel}, a multi-branch scheme is used to extract the spatial information of the input tensor $\bm{X}_{in}$ at different scales. Specifically, we adopt multi-scale convolution kernels with a pyramid structure to obtain the spatial information at different depths and scales. Subsequently, for different branches, each channel-wise feature map $\mathcal{F}_i \in \mathbb{R}^{\bar{C} \times H \times W}$ with the same channel dimension $\bar{C}=C/S$ is given by
\begin{align}
	\mathcal{F}_i = \mathrm{Conv}(k_i,k_i)(\bm{X}_{in}), \; i = \{1,2,...,S\},
\end{align}
 where the \textit{i}-th kernel size $k_i = 2\times i+1$. The whole feature map $\mathcal{F} \in \mathbb{R}^{C \times H \times W}$ can be gained by a concatenation operation as
 \begin{align}
 	\mathcal{F} =\mathcal{F}_1 \oplus \mathcal{F}_2 \oplus...\oplus \mathcal{F}_S, \; i = \{1,2,...,S\}.
 \end{align}

Then, the multi-scale feature map $\mathcal{F}$ is fed into a Gated Squeeze-Excitation (GSE) module to extract the channel attention weight by squeezing the spatial dimension. For the finer channel-wise attention, we split GSE into two branches, respectively using average-pool and max-pool to refine the statistical and discriminative information of objects. Mathematically, each attention weight fragment is written as
     \begin{align}
  	w^{avg/max}_i = \mathrm{GSE}^{avg/max}(\mathcal{F}_i).
  \end{align}

With the corresponding attention fragments concatenated, the whole multi-scale attention vector $W$ is subsequently aggregated by fusing the two GSE branches in a summation way, which is further normalized by the $\mathrm{Sigmoid}$ operation.
   \begin{align}
  	W = \mathrm{Sigmoid}((w^{avg}_1 \oplus w^{avg}_2 \oplus...\oplus w^{avg}_S )\\ \notag
  	+(w^{max}_1 \oplus w^{max}_2 \oplus...\oplus w^{max}_S ) ),
  \end{align}
Finally, the output $\bm{X}_{out}$ can be calculated by multiplying the normalized multi-scale channel attention $W$ with the feature map $\mathcal{F}$ as $ \bm{X}_{out} = W\odot \mathcal{F}$. Compared with the traditional channel attention, the multi-scale and gating mechanisms of PGSA ensure a smooth flow of spatial and channel features, resulting in a more informative output. \par

\textbf{IC module}: In terms of the IC operation, previous work mainly adopt CNNs that enjoys limited perception of long-range dependencies. To enhance the global representation of $\mathcal{D}_k$, we design a multi-scale split transformer (MSST) that equipped with the Half-Shuffle attention block (HASB) \cite{cai2022degradation}. \par

As shown in Fig. \ref{fig:unet}, the proposed MSST contains two main components. First, similar to the multi-branch scheme of PGSA, we split the given input $\bm{X}_{k+1}$ into $S$ subsets $\mathcal{S}_i$, $i\in\{1,2,...,S\}$. Subsequently, the 2nd to $S$th subsets are delivered to $3\times3$ depth-wise convolution $d_i$, and the corresponding output is denoted as $\mathcal{B}_i$. Besides, the subset $\mathcal{S}_i(i>2)$ would be fed to $d_i$ after summing with $\mathcal{B}_{i-1}$. Mathematically, $\mathcal{B}_i$ can be written as
   \begin{align}
	\mathcal{B}_i=\left\{
	\begin{aligned}
		&\mathcal{S}_i ,  &{i= 1} ;\\
		&d_i(\mathcal{S}_i) , & {i= 2}; \\
		&d_i(\mathcal{S}_i+\mathcal{B}_{i-1}),& {2<i \leq S}.
	\end{aligned}
	\right.
\end{align}
Then, a whole feature map $\mathcal{B}$ can be obtained by concatenating $\mathcal{B}_{i}$ as
 \begin{align}
	\mathcal{B} =\mathcal{B}_1 \oplus \mathcal{B}_2 \oplus...\oplus \mathcal{B}_S, \; i = \{1,2,...,S\}.
\end{align}
The multi-level spatial encoding of our first component endows the map $\mathcal{B}$ with a flexible and adaptive spatial perceptual domain, which paves the way for subsequent attention extraction. After position encoding and normalization, $\mathcal{B}$ is fed into the second component, i.e., a three-layer U-shaped architecture constructed by HASB. Then, the output $\bm{Z}_{k+1}$ is obtained via three stages: encoder, bottleneck, and decoder. Each layer of the encoder and decoder contains a HASB and an upsampling/downsampling operation. Besides, we add a Channel Squeeze-Excitation (CSE) module to the decoder for information calibration. Note the design of attention mechanism is not within the scope of our IC module. The detailed information about HASB and CSE are available in \cite{cai2022degradation} and \cite{roy2018concurrent}, respectively. \par
Through the collaboration of IF and IC modules, GA-HQS is expected to achieve efficient information throughput and feature enhancement, whose powerful representation capability will be further verified in the following experiments.

\section{Experimental Results}
\textbf{Datasets}: The IXI and single-coil knee FastMRI datasets are utilized to validate the clinical performance of the proposed method. For IXI, the registered T2 images of 578 patients are adopted, all of which are with a size of $256 \times 256 $. For FastMRI, the type of Proton Density-weighted with Fat Suppression (PDFS) of total 588 volumes are selected, whose sizes are of $320 \times 320$. We split each of the two datasets into a ratio of $7:1:2$ for training/validation/testing. In addition, we employ three different k-space undersampling masks, i.e., 1d cartesian (including random and equispaced fraction) and 2d radial. The acceleration rates in all experiments are set to $4\times$ and $8\times$.\par
\textbf{Implementation Details}:  We implement GA-HQS in the Pytorch framework with an NVIDIA GeForce RTX 3090 GPU. Adam is used for model optimization, with the momentum parameters set as (0.9,0.999) and the learning rate of 1e-4. During the training stage, we set epoch as 30, the batch size as 2, the stage number $k=8$, and the split number $S=4$. Peak Signal-to-Noise Ratio (PSNR) and Structural Similarity Index (SSIM) are used as the evaluation criteria for the reconstruction results.
\begin{table}[htbp]
	\setlength
	\tabcolsep{3.5pt}
	\centering
	\caption{Comparisons between GA-HQS and SOTA methods using three masks on IXI and FastMRI datasets. Params, PSNR, and SSIM are reported.  }
	\scalebox{0.6}{
		\begin{tabular}{cccccccccc}
			\toprule
			&		\multicolumn{3}{c}{Params} & 1.137M  & 7.756M & 1.189M &2.620M & 1.233M& 4.420M \\
			\cmidrule{1-10}
			&	Acc	& Masks	& Metrics & ModL  & Unet & ISTA-Net &MICCAN &  HQS-Net& GA-HQS  \\
			\midrule
			
			\multirow{14}{*}{IXI}	&	\multirow{7}{*}{$4 \times$}& \multirow{2}{*}{\textit{radial}}& PSNR $\uparrow$	&42.542
			&34.053 &43.233 &41.634 & 35.139 &	\textbf{45.286} \\
			&	& &SSIM $\uparrow$	& 0.991 & 0.967 &0.992 & 0.991 &	0.969  & \textbf{0.994}   \\
			\cmidrule{3-10}
			& & \multirow{2}{*}{\textit{random}}& PSNR $\uparrow$	& 33.804  &31.276 &33.612 &35.433 &31.488  &	\textbf{37.077} \\
			&		& &SSIM $\uparrow$	& 0.962 &0.954 &0.960 & 0.973 &	0.948  &\textbf{ 0.981}  \\
			\cmidrule{3-10}
			&	& \multirow{2}{*}{\makecell[c]{\textit{equispaced }\\ \textit{fraction}}}& PSNR $\uparrow$	&30.450   &30.22 &29.890 &33.498 & 30.344 &\textbf{34.573}	 \\
			&	& &SSIM $\uparrow$	& 0.948 & 0.946 & 0.921 & 0.965  &0.942 & \textbf{0.974}  \\
			\cmidrule{2-10}
			&		\multirow{7}{*}{$8 \times$}& \multirow{2}{*}{\textit{radial}}& PSNR $\uparrow$	& 31.266 &29.752 & 31.976&32.830 & 29.017 &\textbf{34.129}	 \\
			&	& &SSIM $\uparrow$	& 0.939 &0.935 &0.941 &0.959  &0.921 &\textbf{0.968 }  \\
			\cmidrule{3-10}
			&	& \multirow{2}{*}{\textit{random}}& PSNR $\uparrow$	& 29.206 &29.057 &28.329 &31.687 & 28.697 & \textbf{33.580}	 \\
			&	& & SSIM $\uparrow$	& 0.938 &0.932 &0.899 &0.950  &0.923	  & \textbf{0.962}  \\
			\cmidrule{3-10}
			&	& \multirow{2}{*}{\makecell[c]{\textit{equispaced} \\ \textit{fraction}}}& PSNR $\uparrow$	&27.721 &27.905 &26.833 &29.987 &27.401  &\textbf{31.287}	 \\
			&	& &SSIM $\uparrow$	& 0.918 &0.9220 &0.886 &0.940  &0.909  &\textbf{0.955}   \\
			\cmidrule{1-10}
			\multirow{14}{*}{FastMRI}	&	\multirow{7}{*}{$4 \times$}& \multirow{2}{*}{\textit{radial}}& PSNR $\uparrow$	&28.046   &27.264 &28.206 &28.339 &27.822  & \textbf{28.508}	\\
			&	& &SSIM $\uparrow$	& 0.743 &0.745 &0.746 & 0.750 &0.745	  &\textbf{ 0.767}  \\
			\cmidrule{3-10}
			& & \multirow{2}{*}{\textit{random}}& PSNR $\uparrow$	& 29.007   & 27.958 & 29.071 & 29.111 & 28.566  & \textbf{29.983}	 \\
			&		& &SSIM $\uparrow$	&0.831  & 0.811 &0.834  &0.835	 &0.819&  \textbf{0.848}   \\
			\cmidrule{3-10}
			&	& \multirow{2}{*}{\makecell[c]{\textit{equispaced} \\ \textit{fraction}}}& PSNR $\uparrow$	& 28.146   & 27.264&28.206 &28.339 &27.822  &\textbf{28.508}	 \\
			&	& &SSIM $\uparrow$	&0.793  &0.780 &0.797 &0.802  &0.787	  &\textbf{0.807}   \\
			\cmidrule{2-10}
			&		\multirow{7}{*}{$8 \times$}& \multirow{2}{*}{\textit{radial}}& PSNR $\uparrow$	&30.145   &28.689 & 30.282&30.205 &29.319  &	\textbf{30.821} \\
			&	& &SSIM $\uparrow$	& 0.863  &0.830 &0.867 &0.865 &0.839 & \textbf{0.873}    \\
			\cmidrule{3-10}
			&	& \multirow{2}{*}{\textit{random}}& PSNR $\uparrow$	&26.983    &26.375 & 27.028& 27.374& 26.636 &\textbf{27.891}	 \\
			&	& &SSIM $\uparrow$	& 0.761 &0.754 &0.762 &0.771  &0.754	  &  \textbf{0.791} \\
			\cmidrule{3-10}
			&	& \multirow{2}{*}{\makecell[c]{\textit{equispaced} \\ \textit{fraction}}}& PSNR $\uparrow$ &27.114 &26.424 &27.259 &27.414 &27.667  &\textbf{27.818}	 \\
			&	& &SSIM $\uparrow$	&0.740  &0.723 &0.741 &0.745  &0.720  & \textbf{0.754}  \\
			\bottomrule
		\end{tabular}%
	}
	\label{table:index}
	\vspace{-10pt}
\end{table}%
\subsection{Comparison with the state-of-the-arts}
\textbf{Quantitative Comparisons}: Table \ref{table:index} compares GA-HQS with five SOTA methods in terms of PSNR and SSIM. The number of parameters for each model are also given. The competing methods include two deep learning networks Unet \cite{zbontar2018fastmri}, MICCAN \cite{huang2019mri}, and three deep-unfolding networks MoDL \cite{aggarwal2018modl}, ISTA-Net \cite{zhang2018ista}, and HQS-Net \cite{xin2021learned}. All algorithms are fine-tuned to report the best performance. From this table, the parameters of our GA-HQS are relatively high, outperforming only Unet and lagging behind the other competitors slightly. However, GA-HQS yields the most impressive performance, particularly in the $4\times$ acceleration of IXI's radial mask. The final values of PSNR and SSIM reach 45.286 and 0.994, respectively. In terms of PSNR, our gains over ISTA and MICCAN are respectively 2.053 dB and 3.652 dB, demonstrating the superiority of our proposals. Given these improvements, we believe the slight lag in terms of parameters is acceptable. \par

\textbf{Qualitative Comparisons}: Under $4\times$ accelerated reconstruction using \textit{equispaced fraction} mask, Fig. \ref{fig:visual1} illustrates a comparison of our GA-HQS with other methods on the IXI dataset. The red box manifests a region of interest that has been magnified to show finer details. It can be seen that GA-HQS is able to maintain uniformity and smoothness while producing cleaner content and fewer artifacts. MICCAN has visually comparable quality, yet the error map generated from GA-HQS presents much bluer results (lower error).  This once again demonstrates the superiority of our proposed unfolding framework.

\begin{table}[htbp]
	\setlength
	\tabcolsep{3.5pt}
	\centering
	\caption{Ablation on the main components of GA-HQS .  }
	\scalebox{0.8}{
		\begin{tabular}{ccccc}
			\toprule
			& 	 baseline-1  & baseline-2  & baseline-3  & GA-HQS   \\
			\midrule
			Acceleration	& \ding{55}& \checkmark & \checkmark &\checkmark       \\	
			PGSA	&\checkmark & \ding{55} &\checkmark  & \checkmark     \\	
			MSST 	&\checkmark & \checkmark &\ding{55}  & \checkmark     \\
			\cmidrule{1-5}
			PSNR $\uparrow$&34.783 &34.866 &  35.042& \textbf{35.470}  \\
			SSIM $\uparrow$		&0.973	&0.973 &0.975	& \textbf{0.978}   \\
			\bottomrule
		\end{tabular}%
	}
	\label{table:ablation}
	\vspace{-15pt}
\end{table}%
\subsection{Ablation Studies}
Our ablation analysis focuses on three main components of GA-HQS. All experiments are performed on the IXI dataset using \textit{equispaced fraction} mask for $4\times$ accelerated reconstruction.

\begin{figure}[tbp]
	\centering
	\includegraphics[width=8.8cm,scale=1.00]{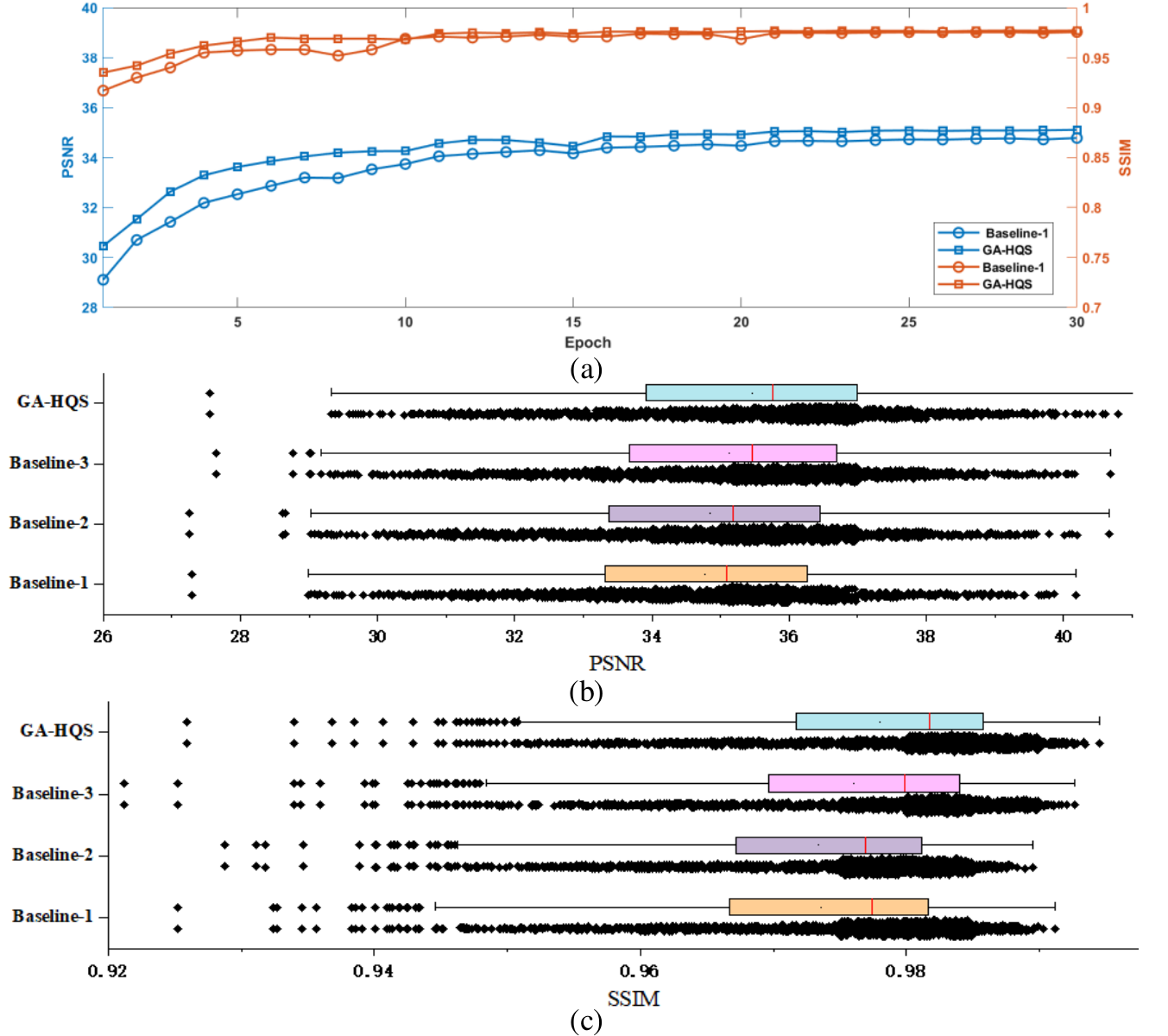}
	\centering
	\caption{Diagram of ablation experiments. (a) Convergence rates of baseline-1 and GA-HQS on PSNR (blue) and SSIM (orange), (b) and (c) are respectively the boxplots of the PSNR and SSIM indices of the test data.}
	\label{fig:ablation1}
	\vspace{-15pt}
\end{figure}

\textbf{Acceleration ablation}. We adopt baseline-1 that is obtained by removing \eqref{fm-hqsc} from \eqref{GA-hqs} to implement the ablation experiment on the acceleration strategy. Based on Fig. \ref{fig:ablation1}(a) and Tab. \ref{table:ablation}, two conclusions can be drawn: firstly, our GA-HQS exhibits a significantly faster convergence rate than baseline-1. Secondly, when the convergence is approaching, GA-HQS yields superior performance. This reveals that incorporating second-order information not only facilitates convergence, but also enhances the overall performance of the algorithm.\par

\textbf{Pyramid gated-squeeze attention (PGSA)}. In order to investigate the role of PGSA in GA-HQS, we adopt baseline-2 that is derived by substituting PGSA with the classical CSE attention. As demonstrated in Fig. \ref{fig:ablation1}(b), (c) and Tab. \ref{table:ablation}, GA-HQS gains 0.604dB improvement over baseline-2. This superiority is derived from PGSA's ability to finely fuse the channel information. \par

\textbf{Multi-scale split transformer (MSST)}. By replacing MSST with Unet, a degraded version of GA-HQS, namely baseline-3, can be formed. The comparisons in Fig. \ref{fig:ablation1}(b), (c) and Tab. \ref{table:ablation} show that the original GA-HQS obtains 0.428dB improvement over baseline-3. Moreover, our results are much more stable. This is mainly attributed to the ability of MSST that captures the non-local dependencies.    \par

\section{Conclusion}
In this paper, we improve upon the previous DUN methods by addressing three problems: the ignorance of the second-order information, the suboptimal information fusion, and the limited ability of CNN on capturing non-local dependencies. Accordingly, we propose a generically unfolding framework based on the accelerated HQS. Moreover, a pyramid-structured channel attention mechanism is incorporated for a fine-grained fusion at the pixel level. Finally, a multi-scale split transformer is designed to capture the global information and enhance the expression ability of the denoising operator. Comprehensive MRI acceleration experiments show that our GA-HQS method outperforms existing state-of-the-art methods. In the near future, we aim to expand more accelerated algorithms and apply them to various reconstruction tasks.

\bibliographystyle{ieeetr}
\begin{spacing}{0.6}
	\bibliography{ref}
\end{spacing}
\end{document}